\documentclass[ pra,showpacs,nofootinbib,draft]{revtex4}
\usepackage{dcolumn,graphicx}

\begin{document} 
 
\title
{Theoretical Study of Sodium and Potassium Resonance Lines
Pressure Broadened by Helium Atoms} 

\author 
{Cheng Zhu, James F. Babb and Alex Dalgarno}

\affiliation
{ ITAMP, Harvard-Smithsonian Center for Astrophysics, 
60 Garden Street, Cambridge, Massachusetts 02138} 
 
\date{\today} 
 
\begin{abstract} 
 
We perform fully quantum mechanical calculations in the binary approximation
of the emission and absorption profiles of the sodium $3s$-$3p$ and
potassium $4s$-$4p$
resonance lines under the 
influence of a helium perturbing gas. 
We use carefully constructed potential energy surfaces and transition 
dipole moments to compute the emission and absorption coefficients at 
temperatures from $158$ to $3000$~K.
Contributions from quasi-bound states are included. 
The resulting red and blue wing profiles agree well with previous 
theoretical calculations and with experimental measurements.

\end{abstract} 
 
\maketitle

\input{epsf}
      
\section{Introduction}

Pressure broadened line profiles of alkali-metal atoms
are prominent in the
spectra of brown dwarfs
and may be a source of opacity in extrasolar giant
planet atmospheres
\cite{ch1:jaya04,ch1:lodders04,ch1:burrows05,lihe:burrows01,lihe:seager00,lihe:brown01}.
Accurate calculations of these
profiles are important in developing effective
diagnostics of the temperatures,
densities, albedos and composition of the atmospheres
of these objects~\cite{lihe:allard03,lihe:burrows03}.

In a recent paper\cite{nahe:zhu05},
we studied the pressure broadening of the lithium $2s$-$2p$ resonance line.
Here we consider the broadening of
the $3s$-$3p$ line of sodium and 
the $4s$-$4p$ line of potassium
arising from collisions with helium atoms. We carry out full
quantitative quantum-mechanical calculations of the emission and absorption
coefficients in the red and blue wings, based on calculated
potential energy surfaces and transition 
dipole moments.
We include contributions from
quasi-bound states and we allow for the variation of the transition
dipole moment with internuclear distance.
We compare our results with previous theoretical calculations
and experimental measurements.

\section{quantum mechanical calculation}
\label{sec:lihe_theo}

The system we consider here consists of a mixture of alkali-metal atoms in a bath of
helium atoms.
If the gas densities are low enough that 
only binary collisions occur, 
the problem can be reduced to the radiation of  
temporarily formed AHe (A$=$Na, K) molecules. 
The broadened $ns$-$np$ 
atomic line corresponds to  
transitions between the excited 
A$^2\Pi$ and B$^2\Sigma$ electronic states 
of the AHe molecule and the ground X$^2\Sigma$ state. 
The X$^2\Sigma$ and B$^2\Sigma$ states have no bound states 
and the A$^2\Pi$ state has a shallow well which 
supports a number of bound ro-vibrational levels. 
Thus we consider both  
bound-free and free-free transitions.
The details of the quantum mechanical calculation
of pressure broadening are given in our previous
paper~\cite{nahe:zhu05}.

\section{Potentials and dipole moments} 
\label{sec:lihe_pot}

\begin{figure}
\centerline{\epsfxsize=12.0cm  \epsfbox{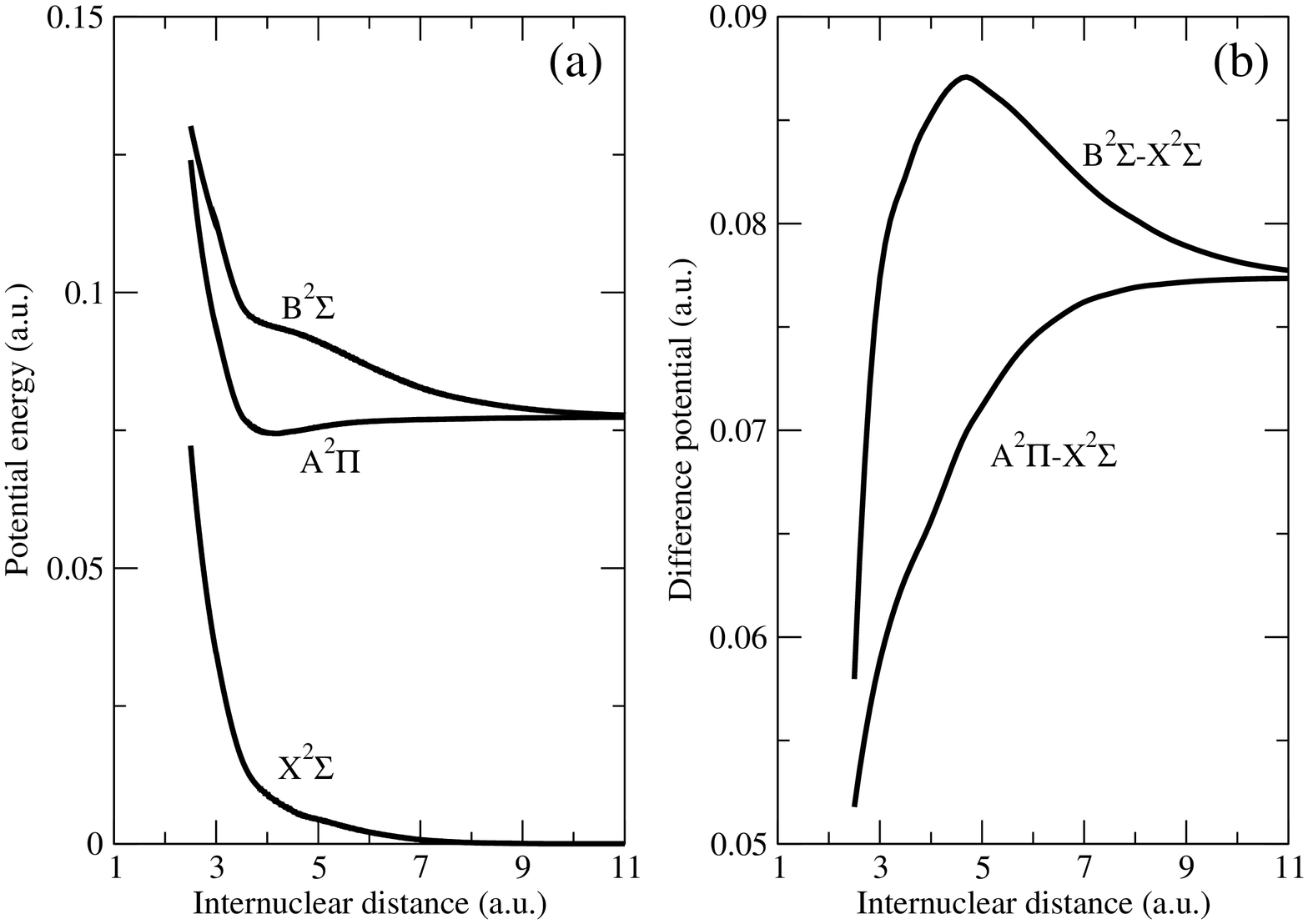}}
\caption{ 
(a) Adopted potentials $V(R)$ for the NaHe $X^2\Sigma$, $A^2\Pi$ and 
$B^2\Sigma$ states. (b) Difference potentials $V_{A^2\Pi}(R)-V_{X^2\Sigma}(R)$ 
and $V_{B^2\Sigma}(R)-V_{X^2\Sigma}(R)$. 
} 
\label{fig:nahe_pot} 
\end{figure}

\begin{figure}
\centerline{\epsfxsize=12.0cm  \epsfbox{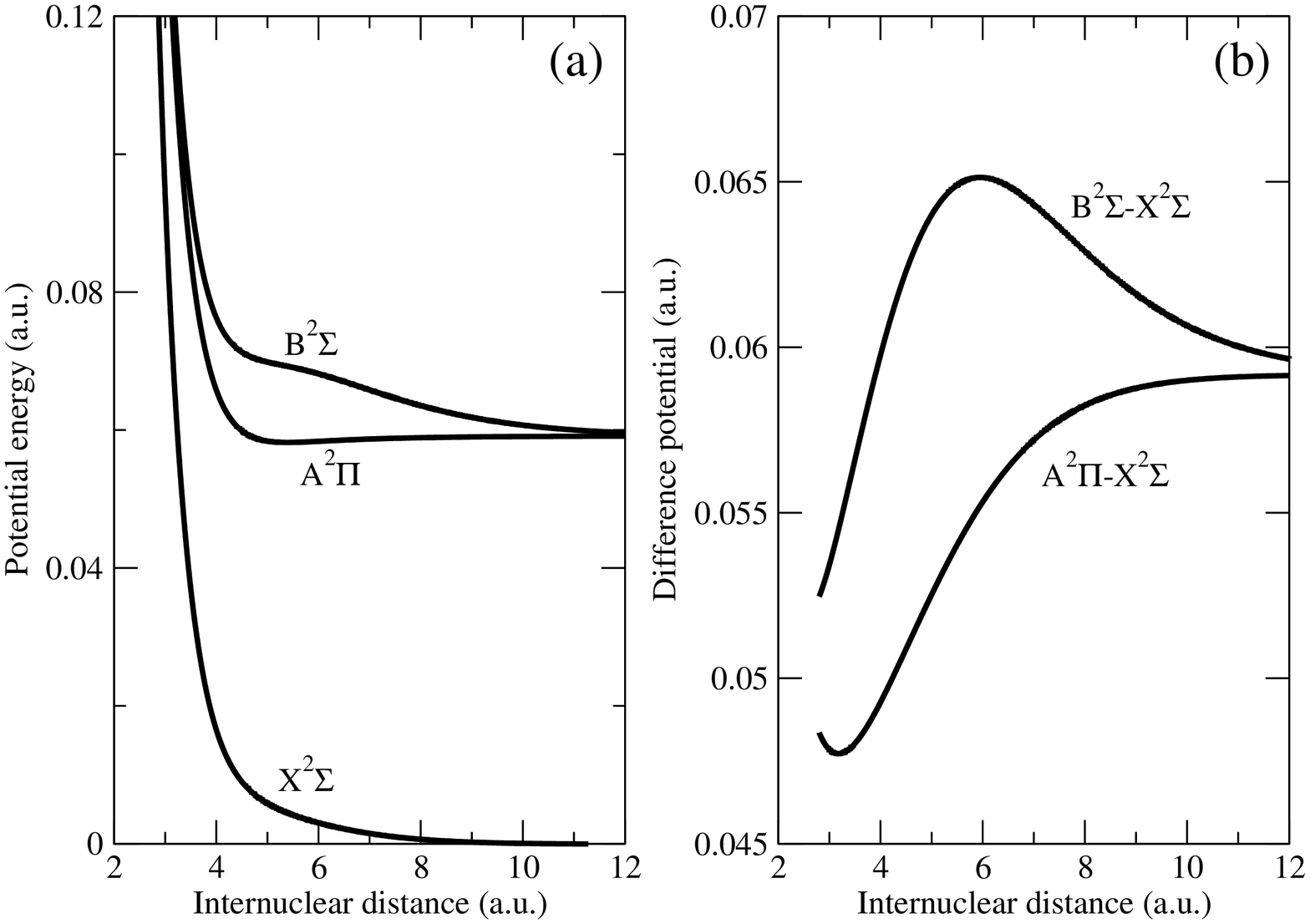}} 
\caption{ 
(a) Adopted potentials $V(R)$ for the KHe $X^2\Sigma$, $A^2\Pi$ and 
$B^2\Sigma$ states. (b) Difference potentials $V_{A^2\Pi}(R)-V_{X^2\Sigma}(R)$ 
and $V_{B^2\Sigma}(R)-V_{X^2\Sigma}(R)$. 
} 
\label{fig:khe_pot} 
\end{figure}

We adopt 
potential energy surfaces computed 
by Theodorakopoulos and Petsalakis~\cite{nahe:theodo93}
for sodium
and by Santra and Kirby~\cite{khe:santra05}
for potassium
for short and intermediate 
internuclear distances.
The $C_6$ coefficients for long range interactions are
taken from the latest theoretical calculations of Zhu
et al.~\cite{nahe:zhu04}.
The potential energy curves of the 
$X^2\Sigma$, $A^2\Pi$ and 
$B^2\Sigma$ states 
are shown in Fig.~\ref{fig:nahe_pot} for NaHe
and Fig.~\ref{fig:khe_pot} for KHe.
The wavelengths $\lambda (R)$
corresponding to the energy
differences $V_{A^2\Pi}(R)-V_{X^2\Sigma}(R)$ 
and $V_{B^2\Sigma}(R)-V_{X^2\Sigma}(R)$ are presented in 
Figs.~\ref{fig:nahe_wav} and \ref{fig:khe_wav}
for NaHe and KHe respectively.
Extremes in the energy differences as a function
of $R$ lead to the presence of satellite features in
the emission and absorption spectra.
The well-depth of the A$^2\Pi$ state of NaHe
is about three times larger than that
of KHe,
but both NaHe and KHe have six bound vibrational levels for
zero angular momentum $J=0$.
Table~\ref{tab:lihe_t1} is a listing of the binding energies of the
vibrational levels of the two molecules.

\begin{figure}
\centerline{\epsfxsize=12.0cm  \epsfbox{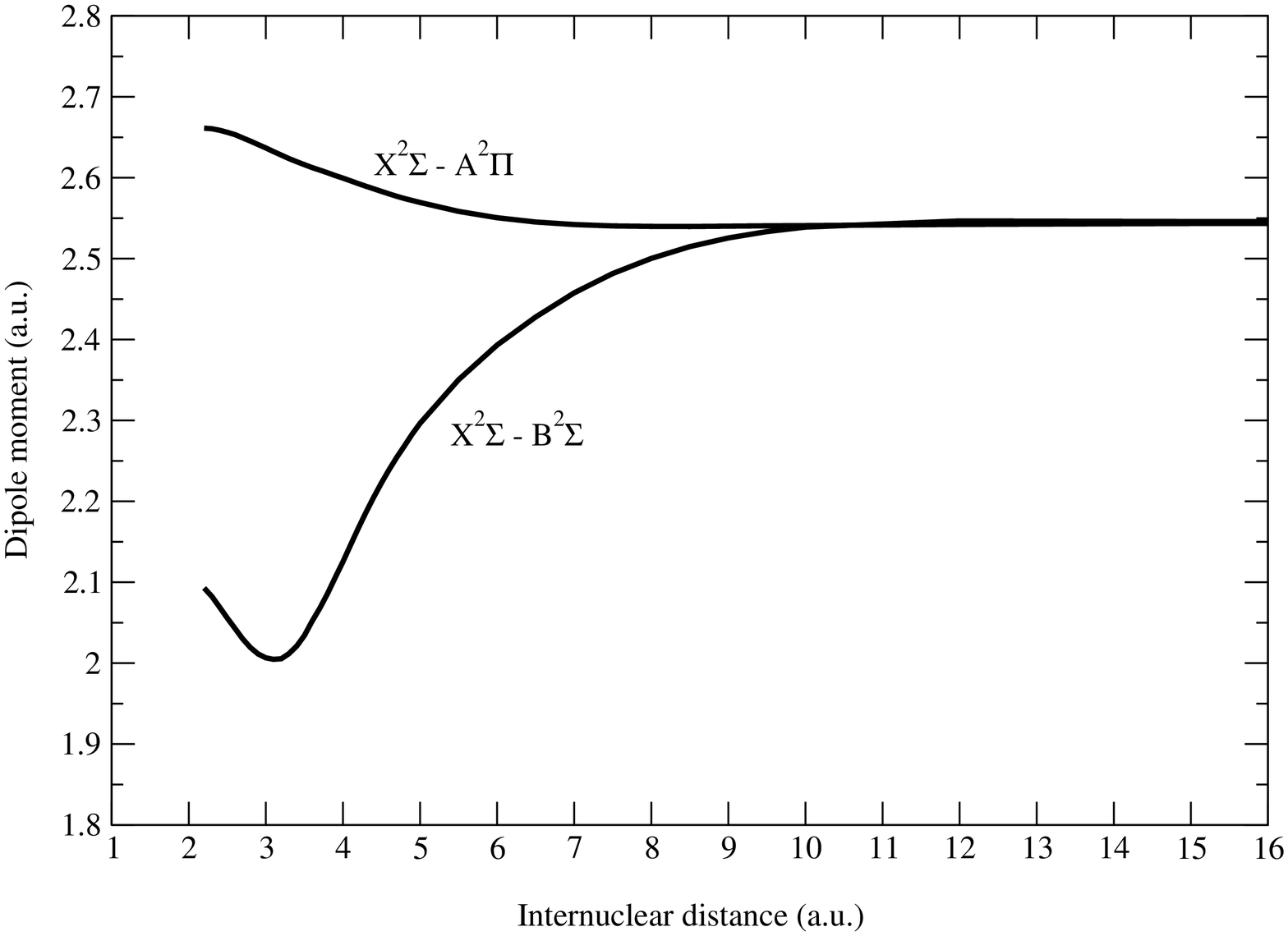}} 
\caption{ 
Adopted dipole moment curves for the NaHe ${X^2\Sigma}-{A^2\Pi}$ 
and ${X^2\Sigma}-{B^2\Sigma}$ transitions. 
} 
\label{fig:nahe_dip} 
\end{figure}

\begin{figure}
\centerline{\epsfxsize=12.0cm  \epsfbox{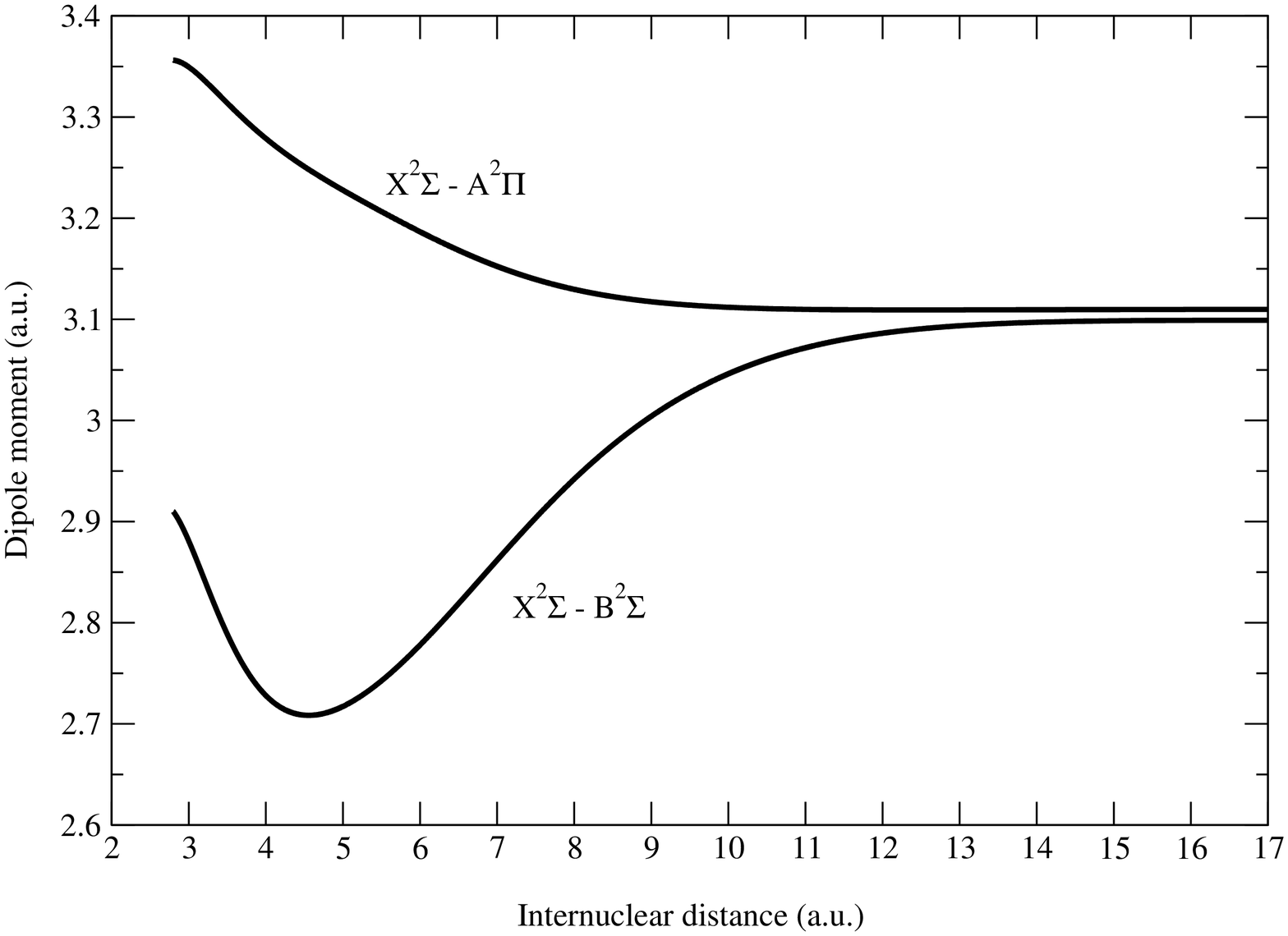}} 
\caption{ 
Adopted dipole moment curves for the KHe ${X^2\Sigma}-{A^2\Pi}$ 
and ${X^2\Sigma}-{B^2\Sigma}$ transitions. 
} 
\label{fig:khe_dip} 
\end{figure}

The adopted ${X^2\Sigma}-{B^2\Sigma}$ and
${X^2\Sigma}-{A^2\Pi}$ 
transition moments~\cite{khe:santra05,nahe:stancil04}
are shown in Fig.~\ref{fig:nahe_dip} for NaHe
and Fig.~\ref{fig:khe_dip} for KHe.

\section{Results and discussions} 
\label{sec:lihe_res}

\begin{figure}
\centerline{\epsfxsize=12.0cm  \epsfbox{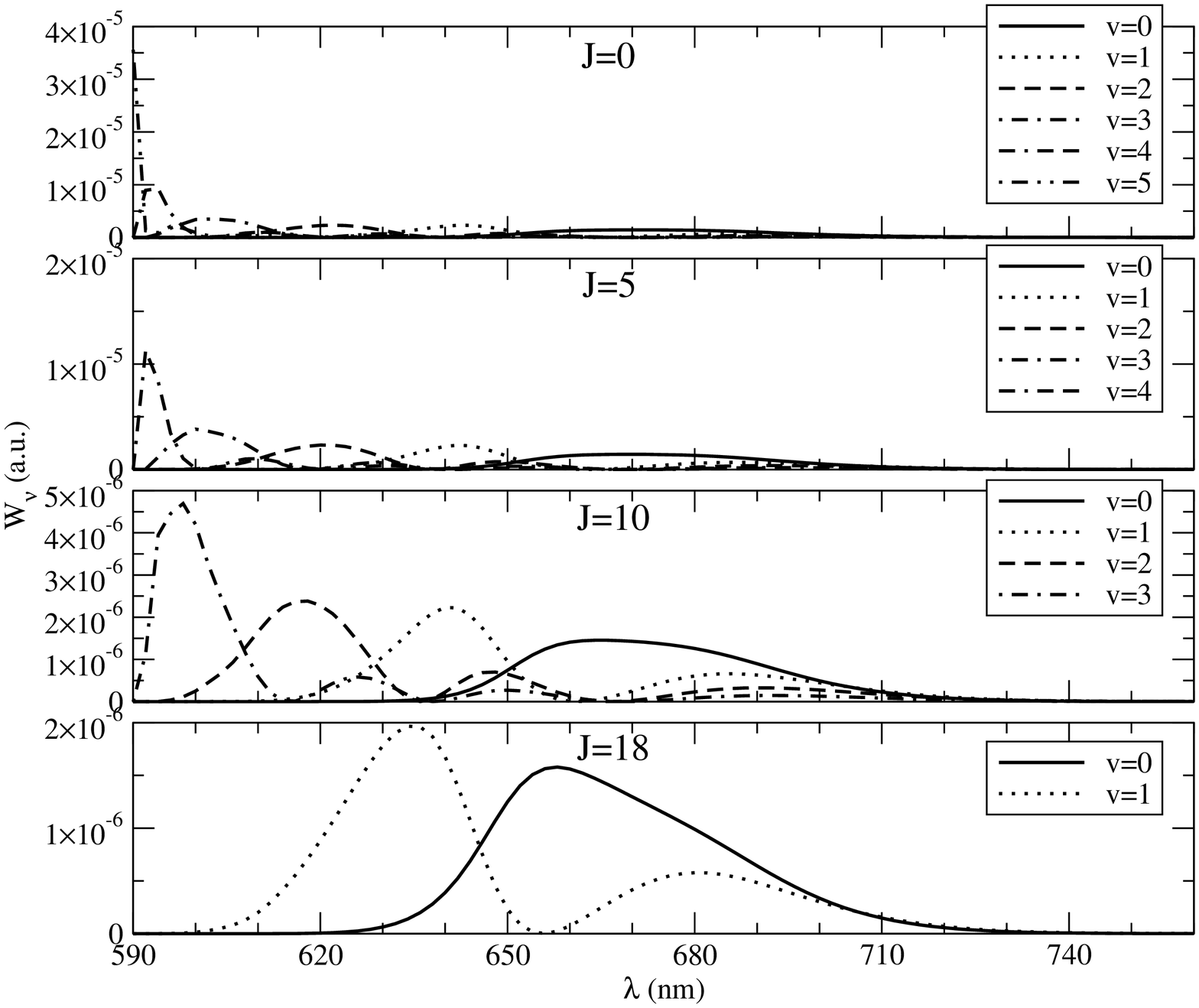}} 
\caption{$W_\nu$ for vibrational levels 
$v=0-5$ for $J=0$ (upper panel), $5$ (upper middle panel),
$10$ (lower middle panel) and $18$ (lower panel) for the NaHe system. 
} 
\label{fig:nahe_ome} 
\end{figure}

\begin{figure}
\centerline{\epsfxsize=12.0cm  \epsfbox{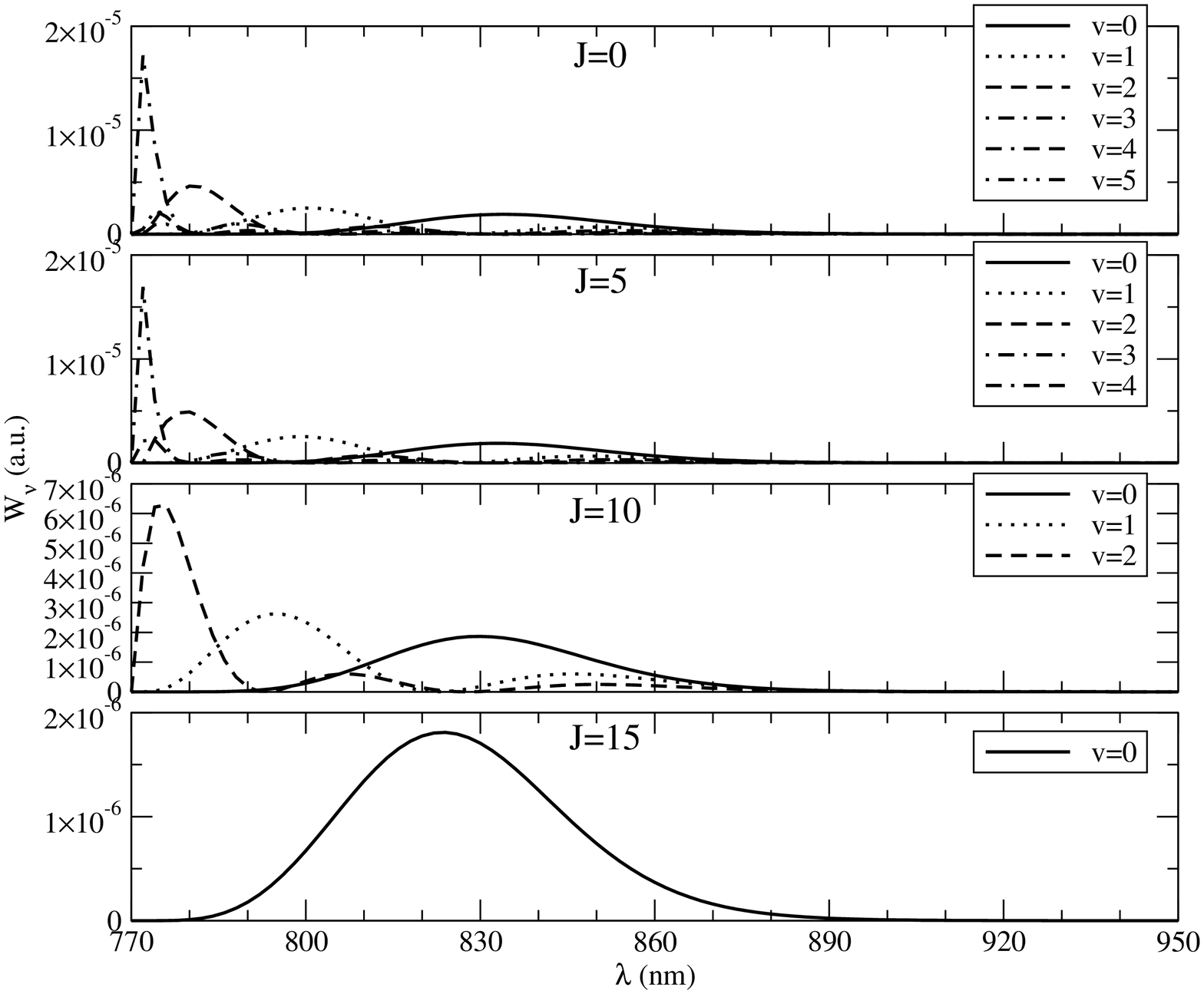}} 
\caption{$W_\nu$ for vibrational levels 
$v=0-5$ for $J=0$ (upper panel), $5$ (upper middle panel),
$10$ (lower middle panel) and $15$ (lower panel) for the KHe system. 
} 
\label{fig:khe_ome} 
\end{figure}

The bound-free emission coefficient is
a weighted sum of the
emission rates $W_\nu$
of the individual ro-vibrational levels of 
the A$^2\Pi$ state~\cite{nahe:zhu05}.
In Figs.~\ref{fig:nahe_ome} and \ref{fig:khe_ome}, we plot  
$W_\nu$  
as a function of wavelength for a number of ro-vibrational 
levels for sodium and potassium, respectively.
The corresponding radiative lifetimes 
are listed in Tables~\ref{tab:nahe_t3} and \ref{tab:khe_t3}.
They do not differ much from 
the lifetimes 
of the excited atoms, $16.3$~ns for Na~\cite{volz96,volz96_2}
and $26.5$~ns for K~\cite{wang97},
into which the A$^2\Pi$ states of the molecules separate
with the exception of the $v=4$ level of KHe.
For the $v=4$ level of KHe there occurs what appears to be a chance
cancellation in the integration of the dipole matrix element.

\begin{figure}
\centerline{\epsfxsize=12.0cm  \epsfbox{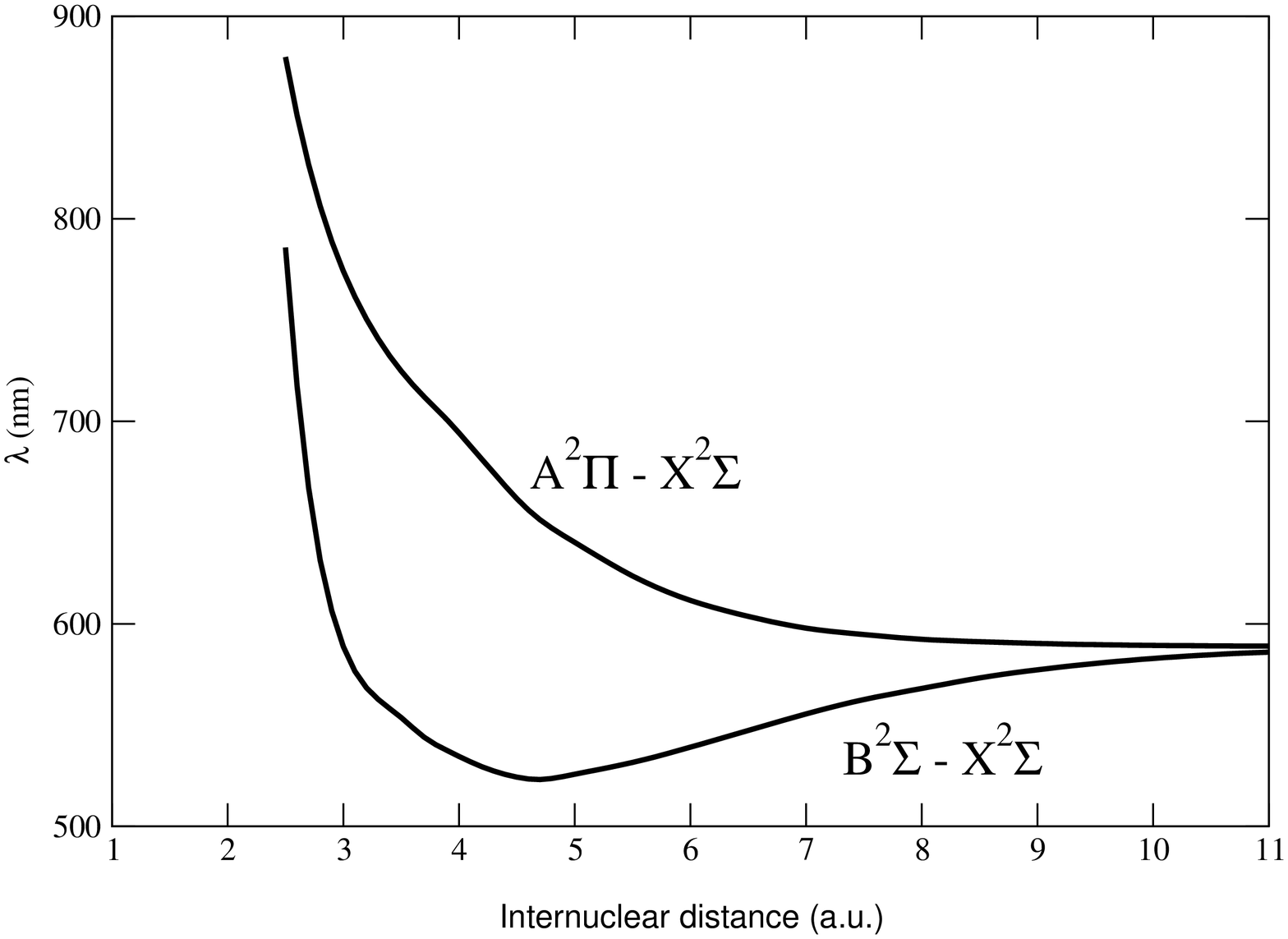}} 
\caption{Wavelengths corresponding to difference potentials
for NaHe A$^2\Pi$-X$^2\Sigma$ and B$^2\Sigma$-X$^2\Sigma$
transitions, respectively.
} 
\label{fig:nahe_wav} 
\end{figure}

\begin{figure}
\centerline{\epsfxsize=12.0cm  \epsfbox{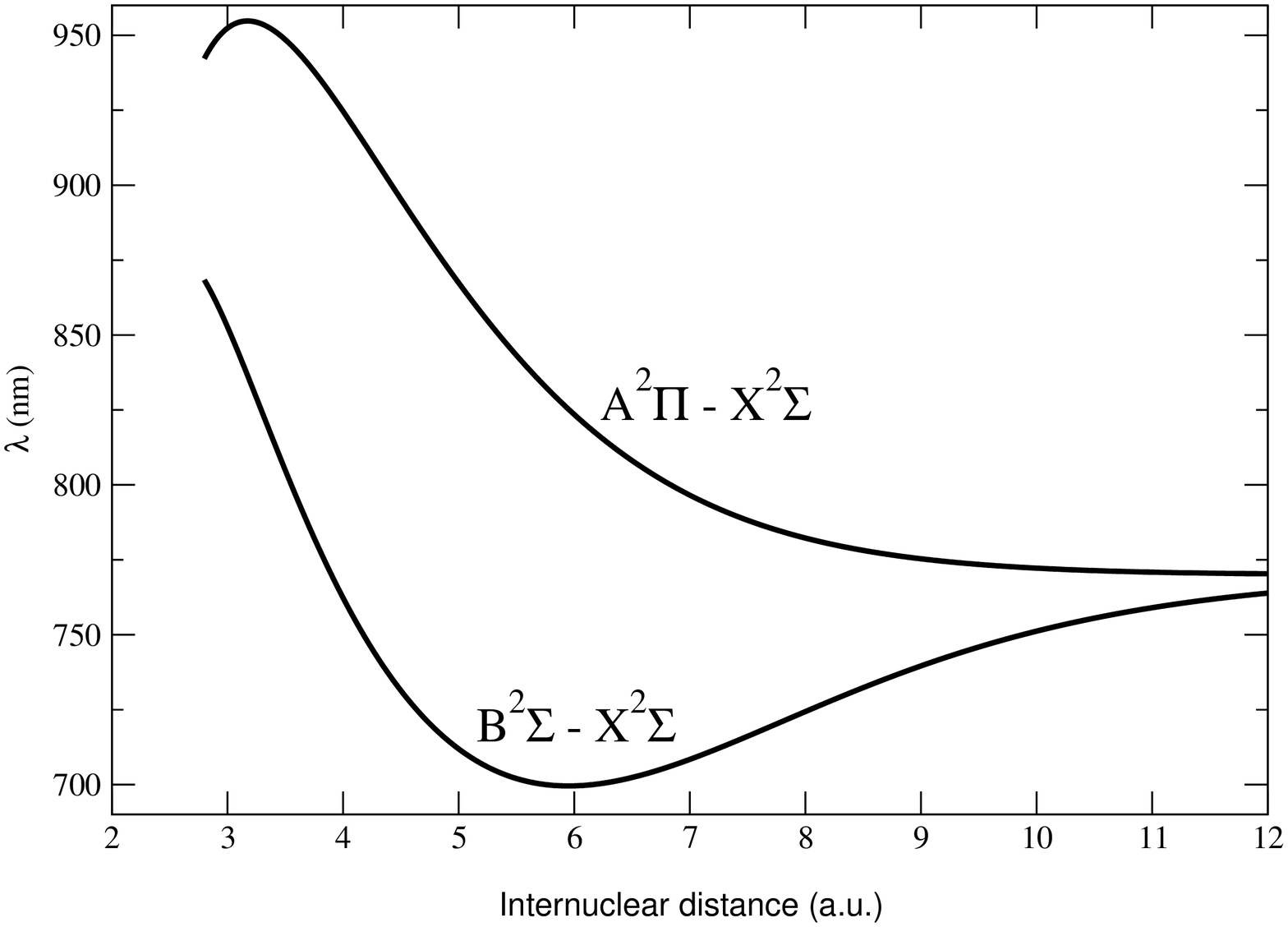}} 
\caption{Wavelengths corresponding to difference potentials
for KHe A$^2\Pi$-X$^2\Sigma$ and B$^2\Sigma$-X$^2\Sigma$
transitions, respectively.
} 
\label{fig:khe_wav} 
\end{figure}

\begin{figure} 
\centerline{\epsfxsize=12.0cm  \epsfbox{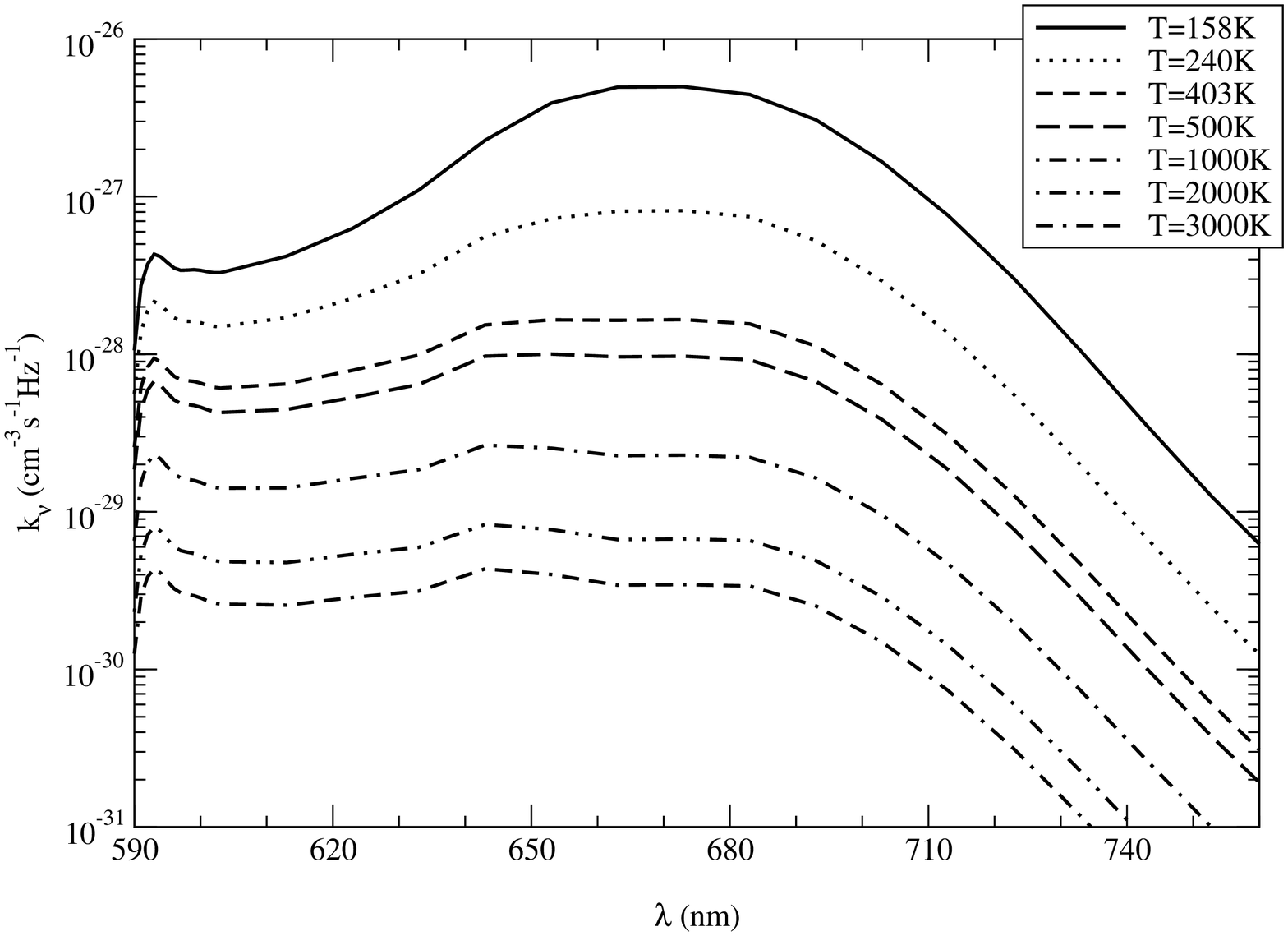}} 
\caption{Contributions of bound-free transitions 
to the total emission coefficients of NaHe at temperatures 
$T=158$, $240$, $403$, $500$, $1000$, $2000$ 
and $3000$K. 
Unit gas densities are used, 
$n_{Na}=n_{He}=1$~cm$^{-3}$. 
} 
\label{fig:nahe_emi1} 
\end{figure}

\begin{figure} 
\centerline{\epsfxsize=12.0cm  \epsfbox{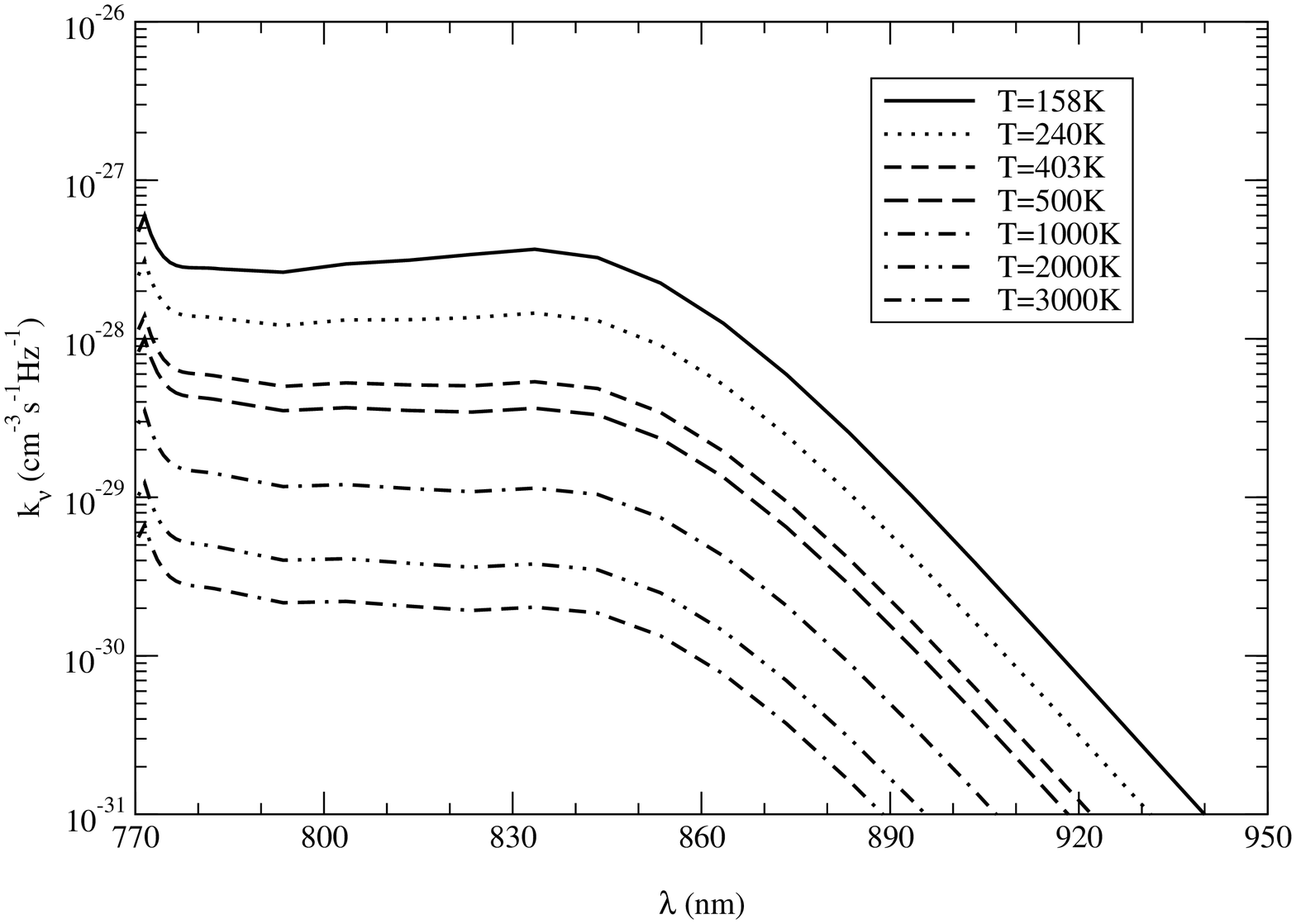}} 
\caption{Contributions of bound-free transitions 
to the total emission coefficients of KHe at temperatures 
$T=158$, $240$, $403$, $500$, $1000$, $2000$ 
and $3000$K. 
Unit gas densities are used, 
$n_{K}=n_{He}=1$~cm$^{-3}$. 
} 
\label{fig:khe_emi1} 
\end{figure}

In Figs.~\ref{fig:nahe_emi1} and \ref{fig:khe_emi1}, we present 
the bound-free contributions to the total emission
coefficients
in units of cm$^{-3}$s$^{-1}$Hz$^{-1}$ for sodium and potassium,
respectively.
The spectra for NaHe and KHe are similar in shape to each other
and to LiHe~\cite{nahe:zhu05} but shifted in wavelength.
At low temperatures there occur broad maxima in the emission
coefficients at around $670$~nm for NaHe and $830$~nm for KHe,
compared to $870$~nm for LiHe.
The emissions are superpositions of the individual ro-vibrational
contributions of Figs.~\ref{fig:nahe_ome} and \ref{fig:khe_ome},
arising from the A$^2\Pi$ states.
For the same temperature, the intensity of the red satellite
is lower for potassium than for sodium.
The emissions decrease rapidly at long wavelengths,
because there are no corresponding values of 
the internuclear distances,
as shown by Fig.~\ref{fig:nahe_wav} for sodium and
Fig.~\ref{fig:khe_wav} for potassium.

\begin{figure} 
\centerline{\epsfxsize=12.0cm  \epsfbox{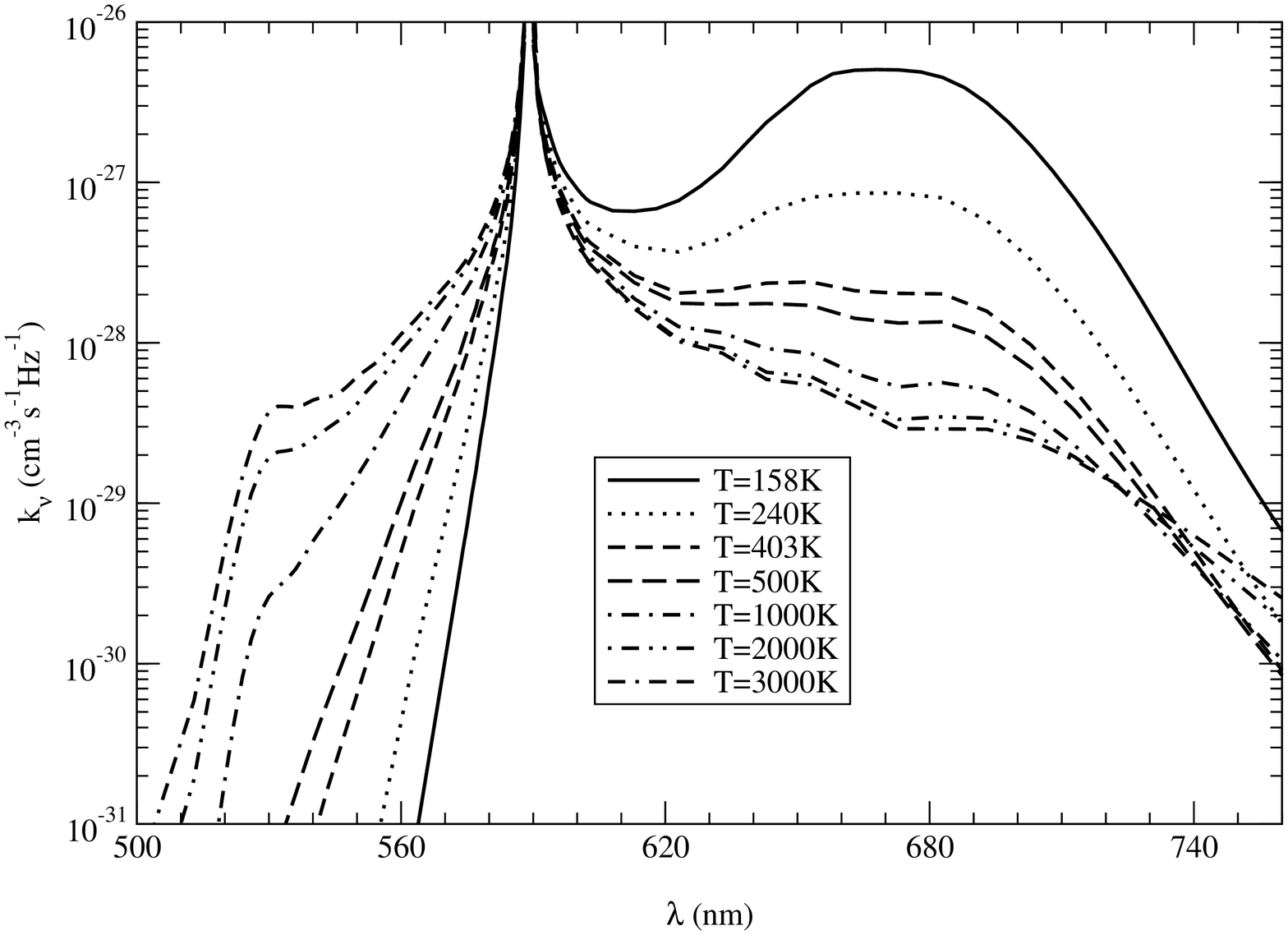}} 
\caption{High pressure emission coefficients of NaHe at temperatures 
$T=158$, $240$, $403$, $500$, $1000$, $2000$ 
and $3000$K. 
Unit gas densities are used, 
$n_{Na}=n_{He}=1$~cm$^{-3}$. 
} 
\label{fig:nahe_emi3} 
\end{figure}

\begin{figure} 
\centerline{\epsfxsize=12.0cm  \epsfbox{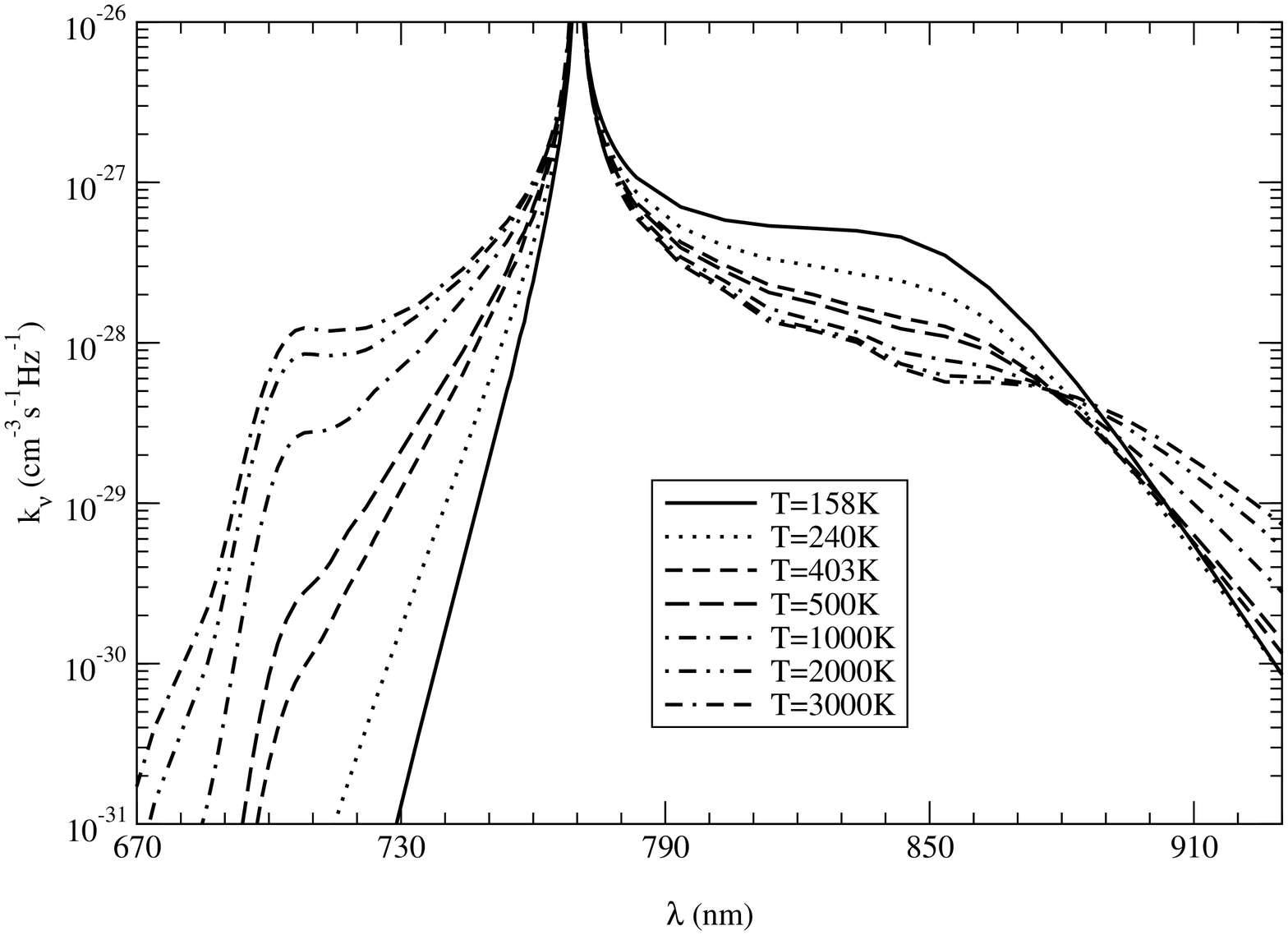}} 
\caption{High pressure emission coefficients of KHe at temperatures 
$T=158$, $240$, $403$, $500$, $1000$, $2000$ 
and $3000$K. 
Unit gas densities are used, 
$n_{K}=n_{He}=1$~cm$^{-3}$. 
} 
\label{fig:khe_emi3} 
\end{figure}

\begin{figure} 
\centerline{\epsfxsize=12.0cm  \epsfbox{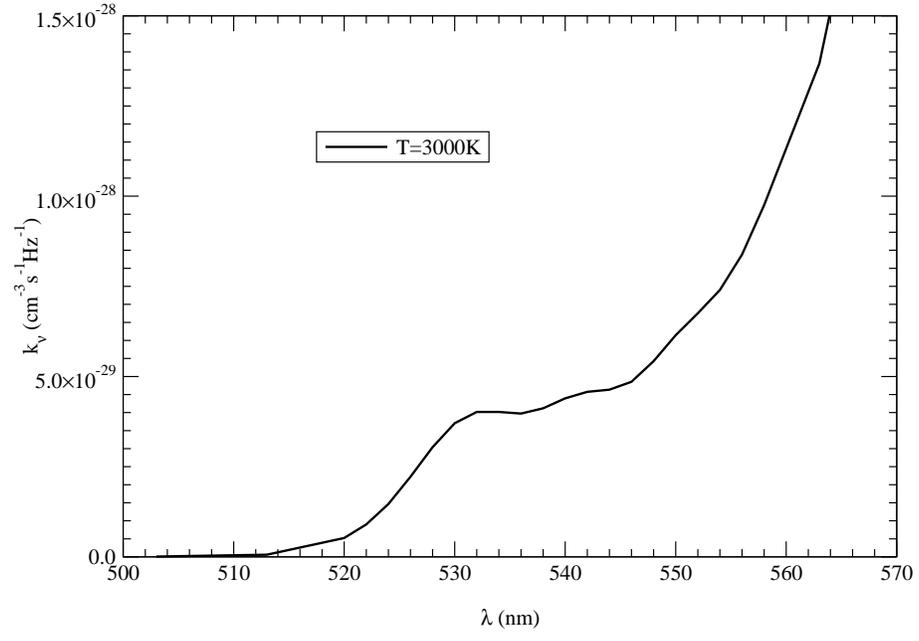}} 
\caption{Emission coefficients of NaHe at temperature
$T=3000$~K plotted on a linear scale.
Unit gas densities are used, 
$n_{Na}=n_{He}=1$~cm$^{-3}$. 
} 
\label{fig:nahe_emi3_linear} 
\end{figure}

\begin{figure} 
\centerline{\epsfxsize=12.0cm  \epsfbox{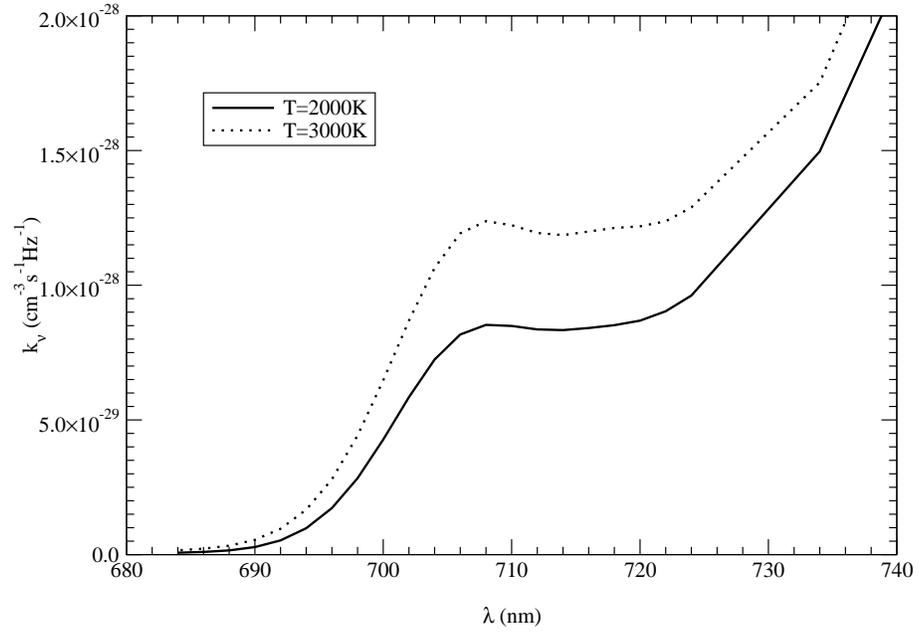}} 
\caption{Emission coefficients of KHe at temperatures 
$T=2000$ and $3000$~K plotted on a linear scale.
Unit gas densities are used, 
$n_{K}=n_{He}=1$~cm$^{-3}$. 
} 
\label{fig:khe_emi3_linear} 
\end{figure}

In the high pressure limit the bound and continuum states are
thermally populated and the emission coefficient is the sum
of the bound-free and free-free coefficients.
In Figs.~\ref{fig:nahe_emi3} and \ref{fig:khe_emi3}, we plot
our calculated emission coefficients
in the high pressure limit 
for temperatures between $158$ and $3000$~K for sodium and
potassium, respectively.
For sodium, the general shape in the red wing is in agreement with
the experimental measurements of
York et al.~\cite{lihe:york75} and
Havey et al.~\cite{lihe:havey80} and
the theoretical calculations of
Pontius and Sando~\cite{lihe:pontius83}
and Theodorakopoulos and Petsalakis~\cite{nahe:theodo93}.
There is no previous calculation or measurement for potassium.
The blue wings
are attributed to
the B$^2\Sigma$-X$^2\Sigma$ transitions, where only
free-free transitions occur and the resulting profiles
decrease rapidly with decreasing wavelength.
A weak satellite appears near $532$ nm at $T=3000$~K for sodium
and near $708$~nm 
at $T\approx 2000$--$3000$~K for potassium,
as expected from a consideration of the $\lambda$--$R$ relations
in Figs.~\ref{fig:nahe_wav} and \ref{fig:khe_wav}.
The satellites are shown
on a linear scale
in Figs.~\ref{fig:nahe_emi3_linear} and \ref{fig:khe_emi3_linear}.

\begin{figure} 
\centerline{\epsfxsize=12.0cm  \epsfbox{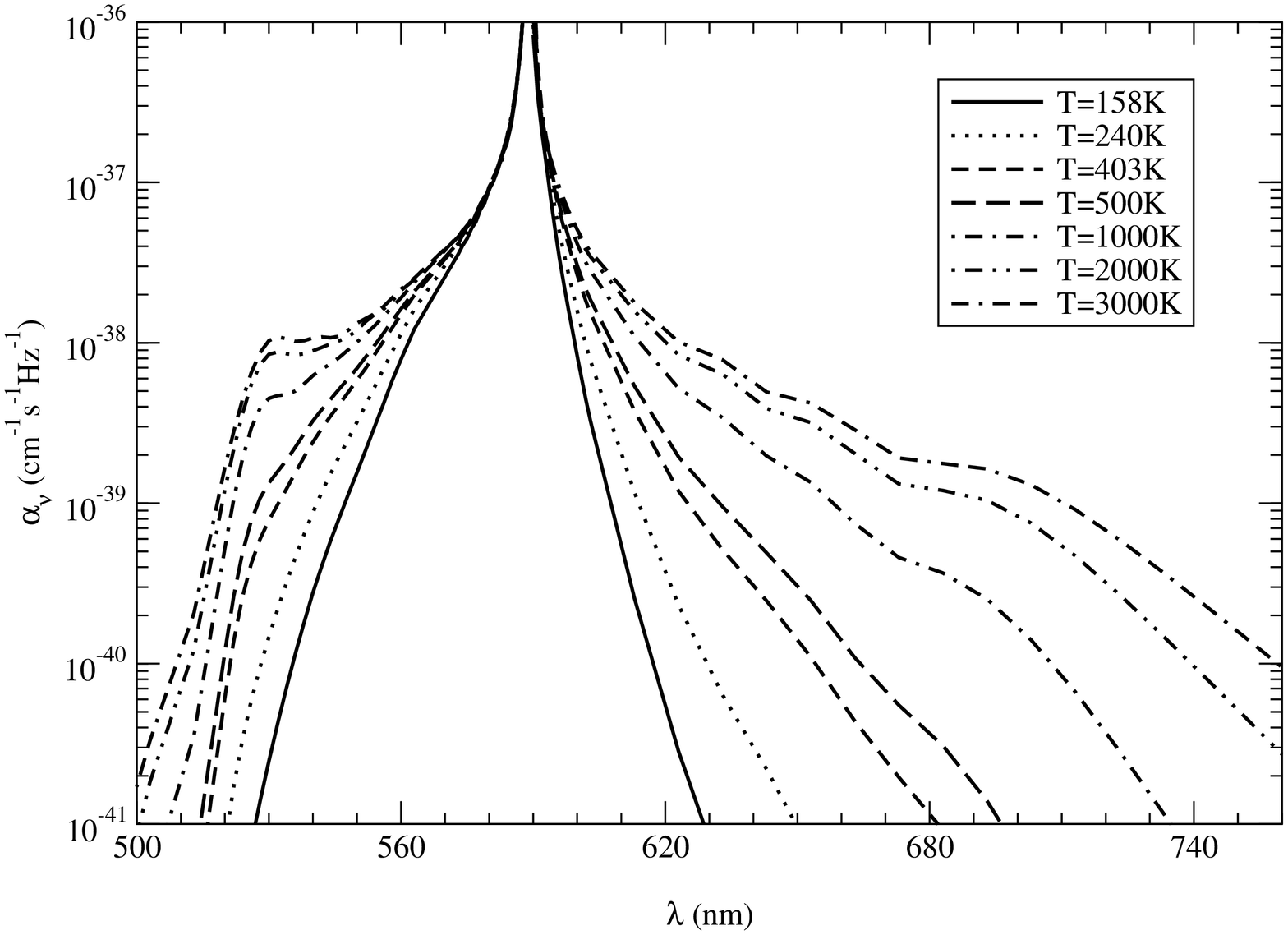}} 
\caption{Absorption coefficients of NaHe at temperatures 
$T=158$, $240$, $403$, $500$, $1000$, $2000$ 
and $3000$K. 
Unit gas densities are used, 
$n_{Na}=n_{He}=1$~cm$^{-3}$. 
} 
\label{fig:nahe_abs} 
\end{figure}

\begin{figure}
\centerline{\epsfxsize=12.0cm  \epsfbox{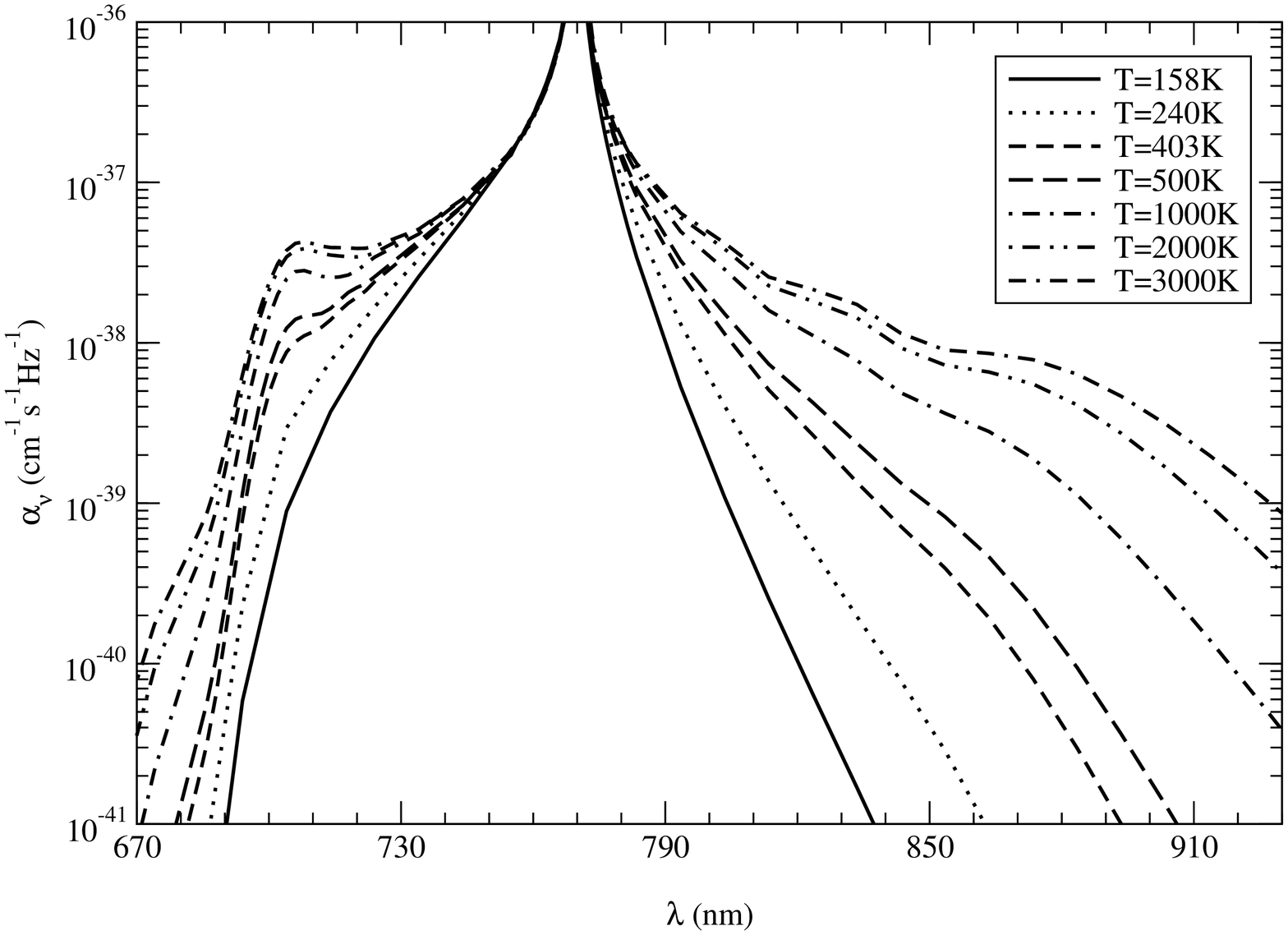}} 
\caption{Absorption coefficients of KHe at temperatures 
$T=158$, $240$, $403$, $500$, $1000$, $2000$ 
and $3000$K. 
Unit gas densities are used, 
$n_{K}=n_{He}=1$~cm$^{-3}$. 
} 
\label{fig:khe_abs} 
\end{figure}

In Figs.~\ref{fig:nahe_abs} and \ref{fig:khe_abs}, we plot our calculated
absorption coefficients for sodium and potassium, respectively. These profiles
decrease steadily
with separation from the position of the resonance line
except for
a blue satellite near $532$~nm
at $T\approx 2000$--$3000$~K for sodium
and near $708$~nm
at $T\approx 1000$--$3000$~K for potassium.
The general shape in the red and blue wings is in
agreement with the theoretical predictions
of Allard et al.~\cite{lihe:allard03}
and Burrows and Volobuyev~\cite{lihe:burrows03}.
Using these calculated absorption profiles,
we can infer from the Reid {\it et al.} (2000) spectrum~\cite{reid00}
of the
L5 dwarf, 2MASSW J1507, that the temperature of this dwarf 
is around $1000$~K.

\section{Conclusions} 
\label{sec:lihe_con}

We have carried out quantum mechanical calculations for 
the emission and absorption spectra 
at wavelengths between
$500$ and $760$~nm for sodium and
between
$670$ and $930$~nm for potassium at temperatures from $158$ to $3000$~K.
For sodium we find a blue satellite near $532$~nm 
for $T=3000$~K and  a red satellite near $670$~nm 
for $T\approx 158$--$240$~K
in the emission spectra,
and a blue satellite  near $532$~nm
for $T\approx 2000$--$3000$~K in the absorption spectra.
For potassium we find a blue satellite near $708$~nm 
for $T\approx 2000$--$3000$~K
and  a red satellite near $830$~nm 
for $T=158$~K
in the emission spectra,
and a blue satellite  near $708$~nm
for $T\approx 1000$--$3000$~K in the absorption spectra.
Our results are in good agreement with previous calculations and
experiment. 
We infer that the temperature of the brown dwarf, 2MASSW J1507,
is around $1000$~K.
 
\acknowledgments 
 
This work was supported in part by 
the NSF through 
a grant for ITAMP to the Smithsonian 
Astrophysical Observatory 
and Harvard University and by NASA under award NAG5-12751.

\clearpage
\newpage

\begin{table}[t]
\begin{center} 
\caption{Vibrational energy levels (cm$^{-1}$) of the A$^2\Pi$ state of 
NaHe and KHe molecules.
}
\label{tab:lihe_t1}
\begin{tabular}{c|cccccc} \hline\hline
$v$                & ~~~~0~~~~ & ~~~~1~~~~ & ~~~~2~~~~ & ~~~~3~~~~
& ~~~~4~~~~ & ~~~~5~~~~  \\
\hline
$E^{v,J=0}$, NaHe & -512.9  & -317.2  & -169.9
& -83.78  & -35.26  & -2.130     \\
$E^{v,J=0}$, KHe & -163.7  & -90.45  & -44.82
& -19.11  & -7.034  & -1.037     \\
\hline\hline 
\end{tabular} 
\end{center}
\end{table}

\clearpage
\newpage

\begin{table}[t]
\begin{center} 
\caption{Lifetimes (ns) of selected ro-vibrational levels $(v,J)$ of
the NaHe A$^2\Pi$ state.
}
\label{tab:nahe_t3}
\begin{tabular}{c|cccccc} \hline\hline
 & ~~~$v=0$~~~ & ~~~$v=1$~~~ & ~~~$v=2$~~~ & ~~~$v=3$~~~
& ~~~$v=4$~~~ & ~~~$v=5$~~~  \\
\hline
$J=0$  & 22.8  & 21.2  & 19.6 &  17.9 & 16.9  &  16.4     \\
$J=5$  & 22.7  & 21.2  & 19.5 &  17.7 & 16.8  &           \\
$J=10$ & 22.5  & 21.0  & 19.1 &  17.3 &       &           \\
$J=18$ & 22.0  & 20.2  &      &       &       &           \\
\hline\hline 
\end{tabular} 
\end{center} 
\end{table} 

\clearpage
\newpage

\begin{table}[t]
\begin{center} 
\caption{Lifetimes (ns) of selected ro-vibrational levels $(v,J)$ of
the KHe A$^2\Pi$ state.
}
\label{tab:khe_t3}
\begin{tabular}{c|cccccc} \hline\hline
 & ~~~$v=0$~~~ & ~~~$v=1$~~~ & ~~~$v=2$~~~ & ~~~$v=3$~~~
& ~~~$v=4$~~~ & ~~~$v=5$~~~  \\
\hline
$J=0$  & 28.3  & 26.6  & 25.2 &  26.2 & 79.8  &  30.0     \\
$J=5$  & 28.2  & 26.4  & 25.1 &  27.1 & 106.9 &           \\
$J=10$ & 27.9  & 26.1  & 24.6 &       &       &           \\
$J=15$ & 27.4  &       &      &       &       &           \\
\hline\hline 
\end{tabular} 
\end{center} 
\end{table}

\clearpage
\newpage

\clearpage
\newpage

\clearpage
\newpage

\clearpage
\newpage

\clearpage
\newpage

\clearpage
\newpage

\clearpage
\newpage

\clearpage
\newpage

\clearpage
\newpage

\clearpage
\newpage

\clearpage
\newpage

\clearpage
\newpage

\clearpage
\newpage

\clearpage
\newpage

\clearpage
\newpage 

\clearpage
\newpage 

\end{document}